\documentclass[twocolumn,floatfix,aps,superscriptaddress,a4paper]{revtex4}
\usepackage{graphicx}
\usepackage{amsmath}
\usepackage{amssymb}
\usepackage{bm}
\usepackage{hyperref}

\usepackage{color}

\begin{document}

\title{Deconfinement of Majorana vortex modes produces a superconducting Landau level}
\author{M. J. Pacholski}
\affiliation{Instituut-Lorentz, Universiteit Leiden, P.O. Box 9506, 2300 RA Leiden, The Netherlands}
\author{G. Lemut}
\affiliation{Instituut-Lorentz, Universiteit Leiden, P.O. Box 9506, 2300 RA Leiden, The Netherlands}
\author{O. Ovdat}
\affiliation{Instituut-Lorentz, Universiteit Leiden, P.O. Box 9506, 2300 RA Leiden, The Netherlands}
\author{\.{I}. Adagideli}
\affiliation{Faculty of Engineering and Natural Sciences, Sabanci University, Orhanli-Tuzla, Istanbul, Turkey}
\affiliation{MESA+ Institute for Nanotechnology, University of Twente, 7500 AE Enschede, The Netherlands}
\author{C. W. J. Beenakker}
\affiliation{Instituut-Lorentz, Universiteit Leiden, P.O. Box 9506, 2300 RA Leiden, The Netherlands}
\date{January 2021}
\begin{abstract}
A spatially oscillating pair potential $\Delta(\bm{r})=\Delta_0 e^{2i\bm{K}\cdot\bm{r}}$ with momentum $K>\Delta_0/\hbar v$ drives a deconfinement transition of the Majorana bound states in the vortex cores of a Fu-Kane heterostructure (a 3D topological insulator with Fermi velocity $v$, on a superconducting substrate with gap $\Delta_0$, in a perpendicular magnetic field). In the deconfined phase at zero chemical potential the Majorana fermions form a dispersionless Landau level, protected by chiral symmetry against broadening due to vortex scattering. The coherent superposition of electrons and holes in the Majorana Landau level is detectable as a local density of states oscillation with wave vector $\sqrt{K^2-(\Delta_0/\hbar v)^2}$. The striped pattern also provides a means to measure the chirality of the Majorana fermions.
\end{abstract}
\maketitle

\emph{Introduction ---}
Deconfinement transitions in physics refer to transitions into a phase where particles can exist as delocalized states, rather than only as bound states. Unlike thermodynamic phase transitions, the deconfinement transition is not associated with a spontaneously broken symmetry but with a change in the momentum space topology of the ground state \cite{Vol07}. A prominent example in superconductors is the appearance of a Fermi surface for Bogoliubov quasiparticles when a superconductor becomes gapless \cite{Agt17,Yua18,Aut20,Lin20}. Such a Bogoliubov Fermi surface has been observed recently \cite{Zhu20}. 

Motivated by these developments we consider here the deconfinement transition for Majorana zero-modes in the vortex core of a topological superconductor. We will demonstrate, analytically and by numerical simulations, that the delocalized phase at zero chemical potential remains a highly degenerate zero-energy level --- a superconducting counterpart of the Majorana Landau level in a Kitaev spin liquid \cite{Rac16,Per17}. Unlike a conventional electronic Landau level, the Majorana Landau level has a non-uniform density profile: quantum interference of the electron and hole components creates spatial oscillations with a wave vector set by the Cooper pair momentum that drives the deconfinement transition.

\begin{figure}[tb]
\centerline{\includegraphics[width=0.9\linewidth]{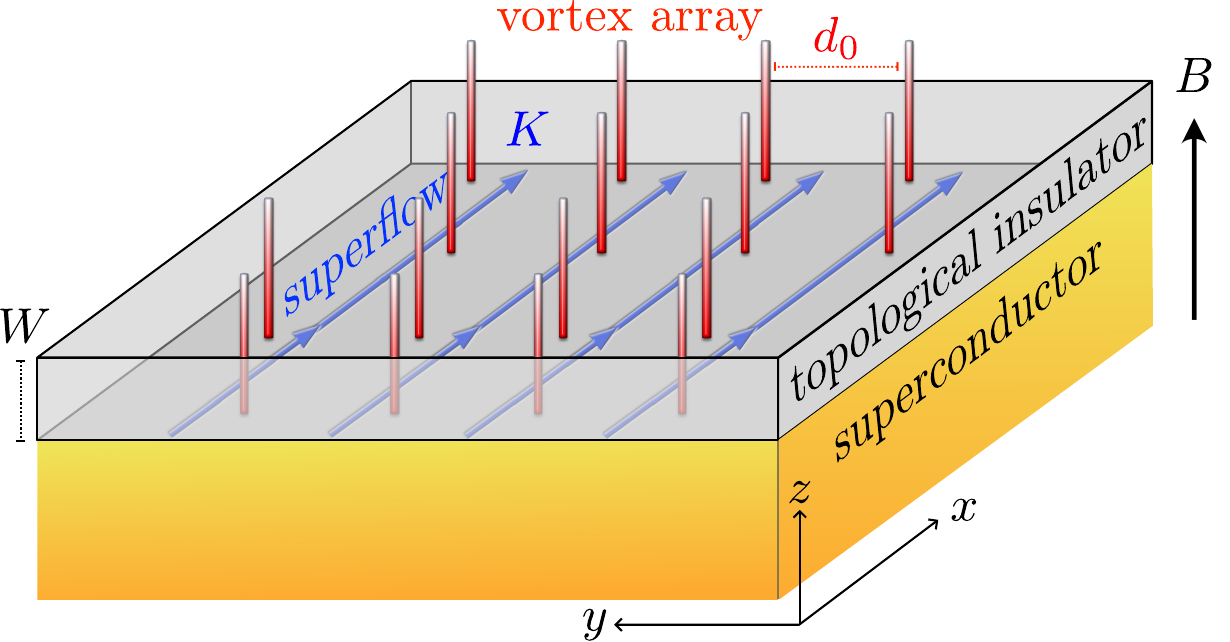}}
\caption{Schematic of the Fu-Kane heterostructure \cite{Fu08}, a topological insulator with induced superconductivity (gap $\Delta_0$) in a perpendicular magnetic field $B$. Vortices (red) bind midgap states known as Majorana zero-modes. Here we study the deconfinement transition in response to an in-plane supercurrent (blue arrows, momentum ${K}$). When $v{K}>\Delta_0$ the zero-modes delocalize into a Majorana Landau level.
}
\label{fig_layout}
\end{figure}

The system of Ref.\ \onlinecite{Zhu20} is shown in Fig.\ \ref{fig_layout}. It is a thin layer of topological insulator deposited on a bulk superconductor, such that the proximity effect induces a pairing gap $\Delta_0$ in the surface states. A superflow with Cooper pair momentum $\bm{K}$ lowers the excitation energy for quasiparticles with velocity $\bm{v}$ by the Doppler shift $\bm{v}\cdot\bm{K}$, closing the gap when $v{K}$ exceeds $\Delta_{0}$. Following Fu and Kane \cite{Fu08}, we add a perpendicular magnetic field $B$ to confine a Majorana zero-mode to the core of each $h/2e$ vortex that penetrates the superconductor. We seek to characterize the deconfined phase that emerges when $v{K}>\Delta_0$.

\emph{Confined phase ---}
To set the stage we first investigate the confined phase for $v{K}<\Delta_0$. Electrons on the two-dimensional (2D) surface of a 3D topological insulator have the Dirac Hamiltonian $v\bm{k}\cdot\bm{\sigma}-\mu$, with $\mu$ the chemical potential, $v$ the energy-independent Fermi velocity, $\bm{k}=(k_x,k_y)$ the momentum operator in the $x$--$y$ surface plane, and $\bm{\sigma}=(\sigma_x,\sigma_y)$ two Pauli spin matrices. (The $2\times 2$ unit matrix $\sigma_0$ is implicit when the Hamiltonian contains a scalar term.) Application of a perpendicular magnetic field $B$ (in the $z$-direction), adds an in-plane vector potential $\bm{A}=(A_x,A_y)$ to the momentum, $\bm{k}\mapsto\bm{k}-e\bm{A}$. The electron charge is $+e$ and for ease of notation we will set $v$ and $\hbar$ both equal to unity in most equations.

The superconducting substrate induces a pair potential $\Delta=\Delta_0 e^{i\phi}$. The phase field $\phi(\bm{r})$ winds by $\pm 2\pi$ around each vortex, at position $\bm{R}_n$, as expressed by
\begin{equation}
\nabla\times\nabla \phi(\bm{r})=\pm 2\pi\hat{z}\textstyle{\sum_{n}}\delta(\bm{r}-\bm{R}_n),\;\;\nabla^2\phi=0.\label{phidef}
\end{equation}
The pair potential couples electrons and holes in the $4\times 4$ Bogoliubov-De Gennes (BdG) Hamiltonian
\begin{equation}
H=\begin{pmatrix}
{K}\sigma_x+(\bm{k}-e\bm{A})\cdot\bm{\sigma}&\Delta_0 e^{i\phi}\\
\Delta_0 e^{-i\phi}&{K}\sigma_x-(\bm{k}+e\bm{A})\cdot\bm{\sigma}
\end{pmatrix},\label{Hdef}
\end{equation}
at zero chemical potential, including a superflow momentum field ${K}\geq 0$ in the $x$-direction \cite{note6}. The superflow can be a screening current in response to a magnetic field in the $y$-direction \cite{Zhu20}, or it can result from an externally imposed flux bias or current bias. The Zeeman energy from an in-plane magnetic field has an equivalent effect \cite{Yua18} (although it was estimated to be negligible relative to the orbital effect of the field in the experiment \cite{Zhu20}).

For $v{K}<\Delta_0$ a pair of Majorana zero-modes will appear in each vortex core, one at the top surface and one at the bottom surface. We consider these separately \cite{note11}. Setting $\Delta(\bm{r})=\Delta_0(r)e^{\pm i\theta}$, in polar coordinates $(r,\theta)$ for a $\pm 2\pi$ phase vortex at the origin, we need to solve the zero-mode equation $H_\pm\Psi_\pm=0$ with
\begin{equation}
H_\pm=\begin{pmatrix}
{K}\sigma_x-(i\nabla+e\bm{A})\cdot\bm{\sigma}&\Delta_0(r) e^{\pm i\theta}\\
\Delta_0(r) e^{\mp i\theta}&{K}\sigma_x+(i\nabla-e\bm{A})\cdot\bm{\sigma}
\end{pmatrix}.
\end{equation}
The pair potential amplitude $\Delta_0(r)$ increases from $0$ at $r=0$ to a value $\Delta_0>0$ when $r$ becomes larger than the superconducting coherence length $\xi_0=\hbar v/\Delta_0$.

When ${K}=0$ this is a familiar calculation \cite{Jac81}, which is readily generalized to $K>0$. The Majorana zero-mode has a definite chirality ${\cal C}$, meaning that its four-component wave function $\Psi_\pm$ is an eigenstate of the chirality operator $\Lambda={\rm diag}\,(1,-1,-1,1)$ with eigenvalue ${\cal C}=\pm 1$. One has $\Psi_+=(i\psi_+,0,0,\psi_+)$, $\Psi_-=(0,i\psi_-,\psi_-,0)$ with \cite{note1}
\begin{subequations}
\label{psismallpssolution}
\begin{align}
&\psi_\pm(\bm{r})=e^{\mp {K}y}e^{\mp\chi(\bm{r})}\exp\left(-\int_0^r \Delta_0(r')\,dr'\right),\\
&\chi(\bm{r})=\frac{e}{2\pi}\int d\bm{r}'\,B(\bm{r}')\ln|\bm{r}-\bm{r}'|.
\end{align}
\end{subequations}
The factor $e^{\mp\chi(\bm{r})}$ is a power law for large $r$, so the zero-mode is confined exponentially to the vortex core as long as ${K}<\Delta_0$. When ${K}>\Delta_0$ the solution \eqref{psismallpssolution} is no longer normalizable, it diverges exponentially along the $y$-axis. This signals a transition into a deconfined phase, which we consider next.

\begin{figure}[tb]
\centerline{\includegraphics[width=1\linewidth]{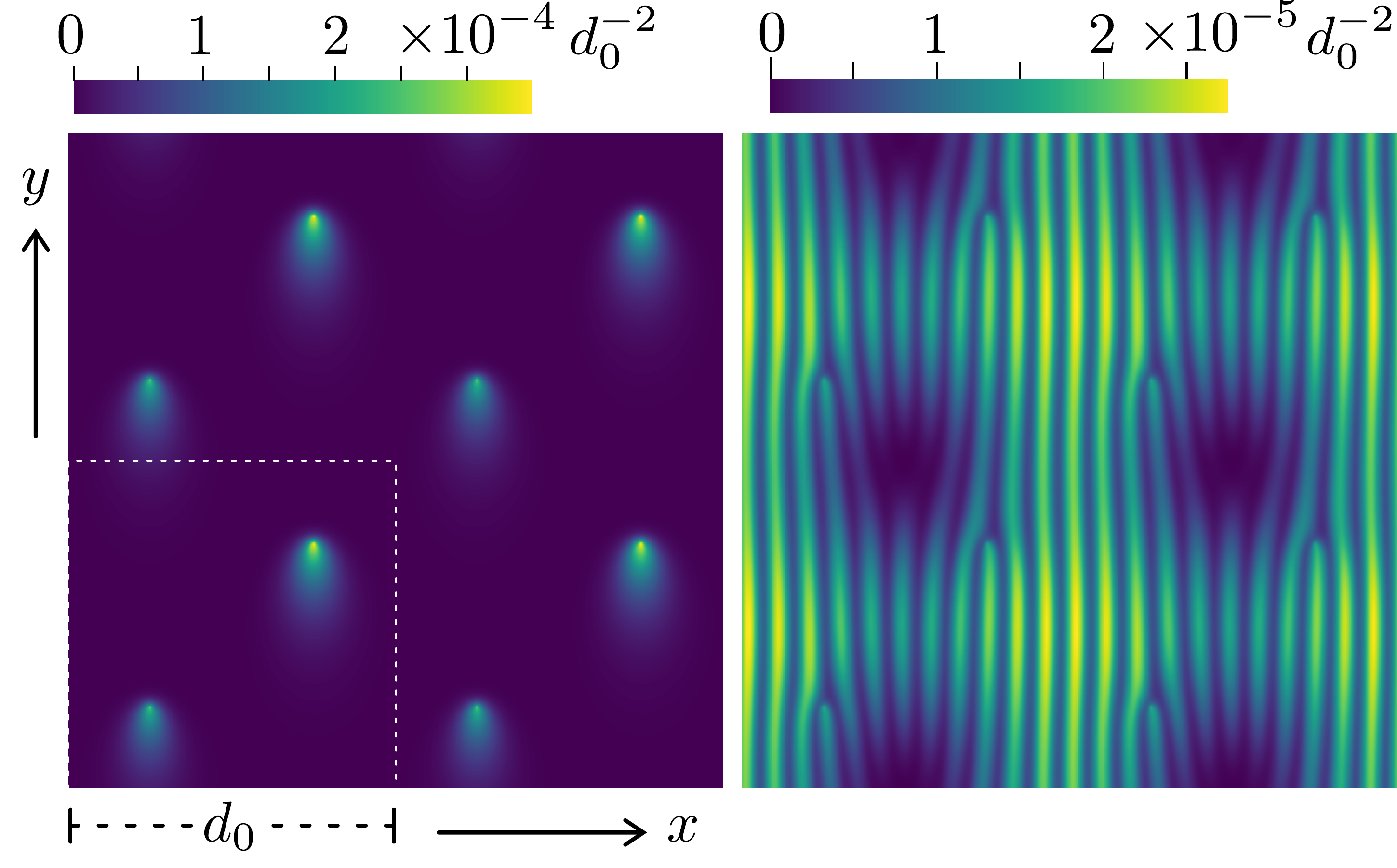}}
\caption{Intensity profile $|\Psi(x,y)|^2$ of a Majorana zero-mode in the vortex lattice \cite{data1}. The left panel shows the confined phase ($K<\Delta_0$), the right panel the deconfined phase ($K>\Delta_0$). The dotted square indicates the unit cell containing a pair of $h/2e$ vortices. These plots are for Majorana fermions of positive chirality, for negative chirality the density profile is inverted $y\mapsto -y$.
}
\label{fig_simulation1}
\end{figure}

\emph{Deconfined phase ---}
In Fig.\ \ref{fig_simulation1} we show results from a numerical simulation of the deconfinement transition for the model Hamiltonian described below. The left panel shows zero-modes confined to a pair of vortex cores for ${K}<\Delta_0$, the right panel shows the deconfined state for ${K}>\Delta_0$. The decay $|\Psi|\propto e^{-{K}y}e^{-\Delta_0 r}$ in the confined phase is anisotropic, with a decay rate $\Delta_0$ along the $x$-axis and two different decay rates $\Delta_0\pm {K}$ in the $\pm y$-direction. The direction into which the zero-mode decays more slowly is set by the chirality \cite{note9}: Fig.\ \ref{fig_simulation1} shows ${\cal C}=+1$ with a slow decay in the $-y$ direction, for ${\cal C}=-1$ the slow decay is in the $+y$ direction.

In the deconfined phase the zero-mode density profile has a pronounced periodic modulation in the $x$-direction, parallel to the superflow, with bifuration points at the vortex cores. This striped pattern is unexpected for a Landau level. We present an analytical description.

\emph{Chiral symmetry protected Majorana Landau level ---}
The chiral symmetry of the Hamiltonian \eqref{Hdef} plays a key role in our analysis of the Majorana Landau level, similar to the role it plays for Landau level quantization in graphene \cite{Kat08,Kai09} and in a Weyl superconductor \cite{Pac18}. Chiral symmetry means that $H$ at $\mu=0$ anticommutes with $\Lambda$. The Hamiltonian then becomes block-off-diagonal in the basis of eigenstates of $\Lambda$,
\begin{subequations}
\begin{align}
&U^\dagger HU=\begin{pmatrix}
0&\Xi\\
\Xi^\dagger&0
\end{pmatrix},\;\;U={\footnotesize\begin{pmatrix}
1&0&0&0\\
0&0&1&0\\
0&0&0&1\\
0&1&0&0
\end{pmatrix}},\\
&\Xi=\begin{pmatrix}
k_--eA_-+{K}&\Delta_0 e^{i\phi}\\
\Delta_0 e^{-i\phi}&-k_+-eA_++{K}
\end{pmatrix},\label{Xiequation}
\end{align}
\end{subequations}
where we have abbreviated $k_\pm=k_x\pm ik_y$, $A_\pm=A_x\pm iA_y$. 

A zero-mode is either a wave function $(u,0)$ of positive chirality with $\Xi^\dagger u=0$, or a wave function $(0,u)$ of negative chirality with $\Xi u=0$. The difference between the number of normalizable eigenstates of either chirality is called the index of the Hamiltonian. It is topologically protected, meaning insensitive to perturbations \cite{Aha79}. 

Vortices are strong scatterers \cite{Mel99}, completely obscuring the Landau level quantization in a nontopological superconductor \cite{Fra00}. Here chiral symmetry ensures that the vortices cannot broaden the zeroth Landau level.

\emph{Helmholtz equation for the Majorana Landau level ---}
Let us focus on the Landau level of positive chirality, described by the equation  $\Xi^\dagger u=0$. This $2\times 2$ matrix differential equation can be simplified by the substitution
\begin{align}
&u(\bm{r})=e^{-{K} y-q(\bm{r})}e^{\tfrac{1}{2}i\phi(\bm{r})\sigma_z}\tilde{u}(\bm{r}),\label{tildevdef}\\
&\text{with}\;\;\partial_x q=-\tfrac{1}{2}\partial_y\phi+eA_y,\;\;\partial_y q=\tfrac{1}{2}\partial_x\phi-eA_x,\label{qdef}\\
&\Rightarrow\begin{pmatrix}
-i\partial_x+\partial_y &\Delta_0\\
\Delta_0&i\partial_x+\partial_y
\end{pmatrix}\tilde{u}=0.\label{tildeudiff}
\end{align}
The fields $\bm{A}$, $\phi$, and ${K}$ no longer appear explicitly in the differential equation \eqref{tildeudiff} for $\tilde{u}$, but they still determine the solution by the requirements of normalizability and single-valuedness of the zero-mode $u$.

Outside of the vortex core the spatial dependence of the pair potential amplitude $\Delta_0$ may be neglected and one further simplification is possible: Substitution of $\tilde{u}=(f,g)$ gives $g=\Delta_0^{-1}(i\partial_x-\partial_y)f$ and a scalar second-order differential equation for $f$,
\begin{equation}
\nabla^2 f= \Delta_0^2 f.\label{Helmholtz}
\end{equation}
In the context of classical wave equations this is the Helmholtz equation with imaginary wave vector.

Eq.\ \eqref{tildevdef} requires that $\tilde{u}$ and hence $f$ have an exponential envelope $e^{Ky}$ in the $y$-direction. The Helmholtz equation \eqref{Helmholtz} then ties that to a plane wave $\propto e^{\pm iQx}$ in the $x$-direction, with wave vector $Q=\sqrt{K^2-\Delta_0^2}$. This already explains the striped pattern in the numerical simulations of Fig.\ \ref{fig_simulation1}. For a more detailed comparison we proceed to a full solution of the Helmholtz equation.
 
\emph{Analytical solution of the Majorana Landau level wave function ---}
The solutions of Eq.\ \eqref{Helmholtz} for $f$ are constrained by the requirements of normalizability and single-valuedness of $u$. To determine the normalizability constraint we use that the field $q(\bm{r})$ defined in Eq.\ \eqref{qdef} has the integral representation \cite{note2}
\begin{equation}
q(\bm{r})=\frac{1}{2\Phi_0}\int d\bm{r}'\,B(\bm{r}')\ln|\bm{r}-\bm{r}'|-\tfrac{1}{2}\sum_n\ln|\bm{r}-\bm{R}_n|.\label{qintegral}
\end{equation}
We consider ${\cal\ N}$ vortices (each of $+2\pi$ vorticity) in a region $S$ enclosing a flux $\Phi={\cal\ N}\Phi_0$, with $\Phi_0=h/2e$ the superconducting flux quantum \cite{note3}. If we set $B\rightarrow 0$ outside of $S$, the field $q(\bm{r})\rightarrow \tfrac{1}{2} (\Phi/\Phi_0-{\cal N})\ln r=0$ for $r\rightarrow \infty$. In view of Eq.\ \eqref{tildevdef}, normalizability requires that $e^{-{K}y}f$ is square integrable for $r\rightarrow\infty$. Near a vortex core $e^{-q}f\propto |\bm{r}-\bm{R}_n|^{1/2}f$ must be square integrable \cite{note4}.

Concerning the single-valuedness, the factor $e^{i\phi/2}$ in Eq.\ \eqref{tildevdef} introduces a branch cut at each vortex position $\bm{R}_n$, across which the function $f$ should change sign --- to ensure a single-valued $u$. This is a local constraint: branch cuts can be connected pairwise, hence there is no sign change in $f$ on a contour encircling a vortex pair.

We have obtained an exact analytical solution \cite{Appsolution} of the Helmholtz equation in the limit that the separation of a vortex pair goes to zero. We place the two vortices at the origin of a disc of radius $R$, enclosing a flux $h/e$, with zero magnetic field outside of the disc. The envelope function then equals $e^{-q(r)}=r_{\rm min}e^{-r_{\rm min}^2/2R^2}$, with $r_{\rm min}=\min(r,R)$.

The two independent solutions are given by $\tilde{u}=(f_1,f_{0})$ and $\tilde{u}'=\sigma_x \tilde{u}^\ast$, with
\begin{align}
&f_n=2i^{n}e^{-in\theta}\text{K}_n(\Delta_0 r)-\int_{-Q}^{Q}dp\,C_n(p)e^{ixp+y\sqrt{\Delta_0^2+p^2}},\nonumber\\
&C_n(p)=\Delta_0^{-n}(\Delta_0^2+p^2)^{-1/2}\bigl(p-\sqrt{\Delta_0^2+p^2}\bigr)^n.\label{fnCndef}
\end{align}
The vortex pair is at the origin, with $x+iy=re^{i\theta}$, and $\text{K}_n$ is a Bessel function. 


The corresponding zero-modes follow from Eq.\ \eqref{tildevdef},
\begin{equation}\label{uanalytic}
u=e^{-q(r)}e^{-Ky}(e^{i\theta} f_1,e^{-i\theta}f_0),\;\;u'=\sigma_x u^\ast.
\end{equation}
For small $r$ the zero-modes tend to a constant (the factor $1/r$ from $\text{K}_1$ is canceled by the factor $r$ from $e^{-q}$). The large-$r$ asymptotics follows upon an expansion of the integrand around the extremal points $\pm Q$, giving
\begin{equation}
f_n\rightarrow(-1)^n \frac{ e^{Ky}}{\Delta_0^n}\left(\frac{(K+Q)^ne^{-iQx}}{iKx -Qy}-\frac{(K-Q)^ne^{iQx}}{iKx+Qy}\right).\label{largerexp}
\end{equation}

The zero-modes decay as $e^{-Ky}f_n\propto 1/r$ for $r\gg R$, which needs to be regularized for a square-integrable wave function \cite{note10,Sut86,Per06}. In a chain of vortices (spacing $b$), the superposition of the solution \eqref{largerexp} decays exponentially in the direction perpendicular to the chain \cite{Appsolution}. The decay length is $\lambda=bK/Q$ or $\lambda=bQ/K$ for a chain oriented along the $x$-axis or $y$-axis, respectively.

\begin{figure}[tb]
\centerline{\includegraphics[width=0.8\linewidth]{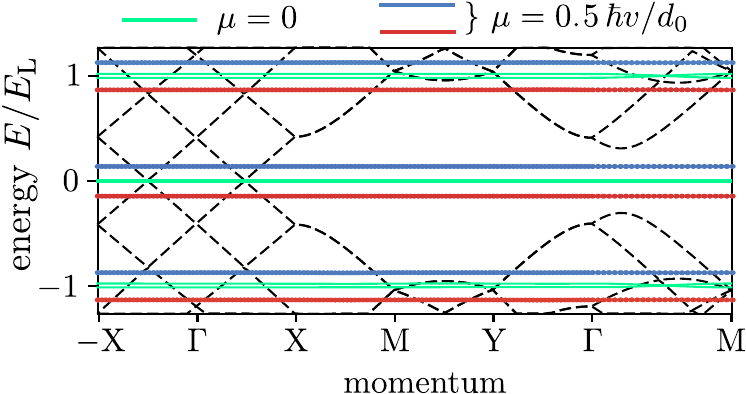}}
\caption{Dispersion relation of the topological superconductor, calculated from the model Hamiltonian \eqref{H02x2} for zero magnetic field (black dashed lines, chemical potential $\mu=0$) and in the presence of the magnetic vortex lattice (colored flat bands at charge $\pm q_{\rm eff}e$, for two values of $\mu$). For both data sets $K=2\Delta_0=20\, \hbar v/d_0$.
}
\label{fig_bandstructure}
\end{figure}

\emph{Numerical simulation ---}
For a numerical study of the deconfinement transition we represent the topological insulator layer by the low-energy Hamiltonian \cite{Sha10,Zha15}
\begin{equation}
\begin{split}
&H_0(\bm{k})=(v/a_0)\textstyle{\sum_{j=x,y}}\sigma_j \sin k_j a_0+ \sigma_z M(k)-\mu,\\
&M(k)=M_0-(M_1/a_0^2)\textstyle{\sum_{j=x,y}}(1-\cos k_ja_0),
\end{split}\label{H02x2}
\end{equation}
in the basis $\Psi=2^{-1/2}(\psi_{\uparrow{\rm upper}}+\psi_{\uparrow{\rm lower}},\psi_{\downarrow{\rm upper}}-\psi_{\downarrow{\rm lower}})$ of spin-up and spin-down states on the upper and lower surfaces \cite{note5}. The atomic lattice constant is $a_0$, the Fermi velocity is $v$, and $\mu$ is the chemical potential. Hybridization of the states on the two surfaces introduces the mass term $M(k)$. We set $M_0=0$, to avoid the opening of a gap at $k=0$ \cite{note11}, but retain a nonzero $M_1=0.2\,a_0v$ in order to eliminate the fermion doubling at $a_0\bm{k}=(\pi,\pi)$. 

\begin{figure}[tb]
\centerline{\includegraphics[width=1\linewidth]{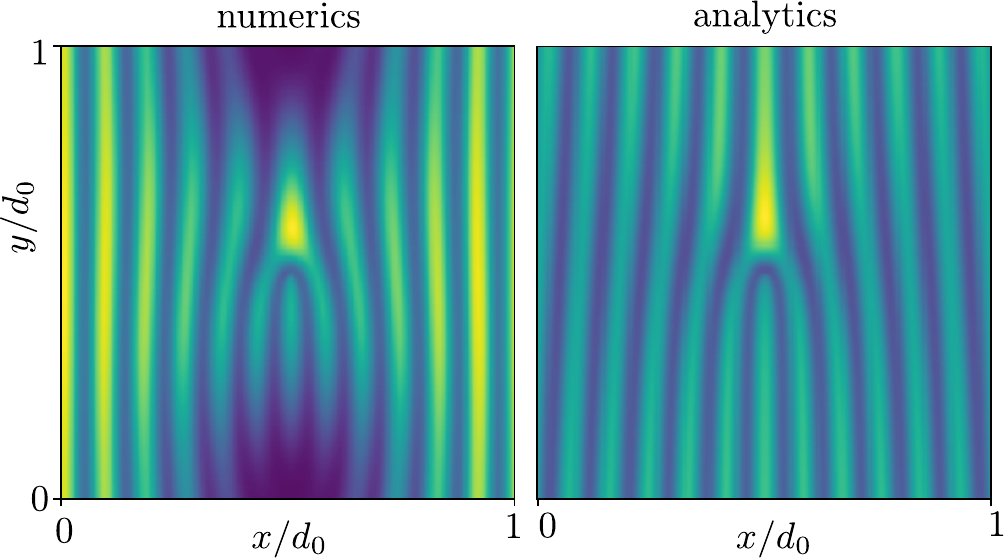}}
\caption{Left panel: Numerically calculated intensity profile $|\Psi(x,y)|^2$ of the zeroth Landau level in a vortex lattice with a pair of $h/2e$ vortices at the center of the unit cell ($K=2\Delta_0=40\, \hbar v/d_0$, $\mu=0$). Right panel: Analytical result from the solution of the Helmholtz equation \eqref{Helmholtz} for a single $h/e$ vortex \cite{note8}.
}
\label{fig_comparison}
\end{figure}

In the corresponding BdG Hamiltonian the electron block $H_0(\bm{k}-e\bm{A}+\bm{K})$ is coupled to the hole block $-H_0(\bm{k}+e\bm{A}-\bm{K})$ by the $s$-wave pair potential $\Delta_0 e^{i\phi}$, which we take the same for both layers. We assume a strong type-II superconductor, for which we can take a uniform magnetic field $B$ and uniform pair potential amplitude $\Delta_0$. The $+2\pi$ vortices are positioned on a square lattice (lattice constant $d_0=302\,a_0$) with two vortices per unit cell.

The spectrum is calculated using the \textit{Kwant} tight-binding code \cite{kwant,Apsimulation}. In Fig.\ \ref{fig_bandstructure} we show the dispersionless Landau levels, both for chemical potential $\mu=0$ and for nonzero $\mu$. The zeroth Landau level has energy $E_0=\pm q_{\rm eff} \mu$, with $q_{\rm eff}e$ the charge expectation value. For the model Hamiltonian \eqref{Hdef} we have \cite{appQeff}
$q_{\rm eff}=Q/K=\sqrt{1-\Delta_0^2/K^2}$. The numerics at $K=2\Delta_0$ gives a value 0.85, within 2\% of $\sqrt{3/4}=0.866$. The first Landau level is expected at energy $E_1 = E_{\rm L}\pm q_{\rm eff}\mu$ with $E_{\rm L}=\sqrt{4\pi q_{\rm eff}}\,\hbar v/d_0$, again in very good agreement with the numerics. Notice that the flatness of the dispersion persists at nonzero $\mu$ --- even though the topological protection due to chiral symmetry \cite{note7} is only rigorously effective at $\mu=0$. 

In Fig.\ \ref{fig_comparison} we compare numerical and analytical results for the case that the two $h/2e$ vortices are both placed at the center of the unit cell. The agreement is quite satisfactory, given the different geometries (a vortex lattice in the numerics, a single $h/e$ vortex in the analytics). 

\emph{Striped local density of states ---}
The striped pattern of the Majorana Landau level is observable by tunneling spectroscopy, which measures the local density of states
\begin{equation}
\rho(\bm{r})=\textstyle{\sum_{\bm{k}}}\bigl[|\psi_{e}(\bm{r})|^2 f'(E_0-eV)+|\psi_{h}(\bm{r})|^2 f'(E_0+eV)\bigr],
\end{equation}
averaged over the 2D magnetic Brillouin zone, $\sum_{\bm{k}}=(2\pi)^{-2}\int dk_x dk_y$, weighted by the derivative of the Fermi function. 
If $E_0$ is much larger than temperature, the sign of the bias voltage $V$ determines whether the electron component $\psi_e$ or the hole component $\psi_h$ contributes, so these can be measured separately.

\begin{figure}[tb]
\centerline{\includegraphics[width=0.7\linewidth]{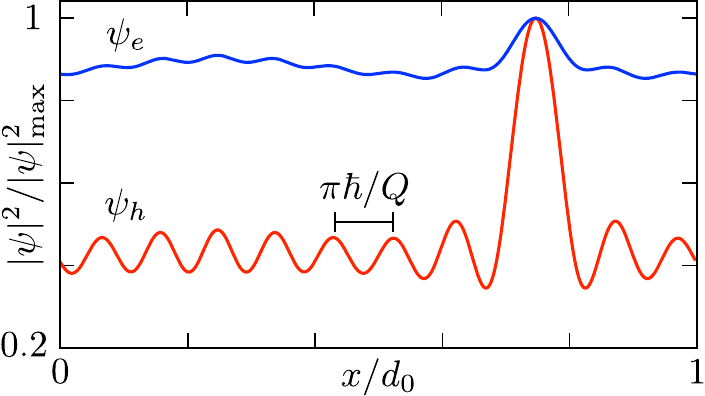}}
\caption{Electron and hole contributions to the local density of states in the zeroth Landau level, along a line parallel to the $x$-axis which passes close through a vortex core at $x=y=3d_0/4$. The curves are plots of $\sum_{\bm{k}}|\psi_{e,h}(x,y)|^2$ normalized to unit peak height at the vortex core. The parameters are $K=2\Delta_0=40\, \hbar v/d_0$, $\mu=0.5\,\hbar v/d_0$. The expected oscillation period of $\pi\hbar/Q=0.091\,d_0$ is indicated.
}
\label{fig_DOS}
\end{figure}

As shown in Fig.\ \ref{fig_DOS}, the oscillations are most pronounced for the hole component when $\mu>0$ (or equivalently the electron component when $\mu<0$). This asymmetry in the tunneling current for $V=\pm E_0$ is an additional experimental signature of the effect.

\emph{Conclusion ---}
Concerning the experimental feasibility, we note that the gap closing due to a superflow has already been observed \cite{Zhu20}, and Majorana vortex lattices in a perpendicular field of $250\,\text{mT}$ have been detected by scanning probes in several experiments \cite{Sun17} --- so by combining these two ingredients the Majorana Landau level should become accessible. The main additional requirement is that the Fermi level is sufficiently small, $\mu<\min(E_{\rm L},\Delta_0)\simeq 1\,\text{meV}$ at $250\,\text{mT}$, to benefit from the protection afforded by chiral symmetry. Experiments \cite{Cho13} where $\mu$ was tuned through the charge neutrality point give confidence that this is feasible. 

The striped interference pattern in the local density of states, with wave number $Q=\sqrt{K^2-(\Delta_0/\hbar v)^2}$ ($\simeq 2\pi/0.2\,\mu\text{m}$ for $K=2\Delta_0/\hbar v$ at typical values of $\Delta_0=1\,\text{meV}$ and $v= 10^5\,\text{m/s}$) should be accessible by scanning probe spectroscopy. Surface defects would themselves introduce Friedel oscillations in the density of states, but the highly directional  pattern that is the hallmark of the Majorana Landau level would stand out.

The Majorana Landau level provides a realization of a flat band with extended wave functions, in which interaction effects are expected to be enhanced due to the quenching of kinetic energy. Interacting Majorana fermions in a Fu-Kane superconductor have been studied by placing vortices in close proximity inside a quantum dot \cite{Pik17}. The deconfinement transition provides a means to open up the system and obtain a fully 2D flat band with widely separated vortices. An intriguing topic for further research is to investigate how the exchange of vortices operates on this highly degenerate manifold. 

\emph{Acknowledgements ---}
We have benefited from discussions with A. R. Akhmerov and A. Donis Vela. This project has received funding from the Netherlands Organization for Scientific Research (NWO/OCW) and from the European Research Council (ERC) under the European Union's Horizon 2020 research and innovation programme.



\appendix

\makeatletter
\renewcommand{\bibnumfmt}[1]{[S#1]}
\renewcommand{\citenumfont}[1]{S#1}

\section{Details of the numerical simulation}
\label{sec_numerics}

\subsection{Tight-binding model}
\label{sec_TB}

The model Hamiltonian we consider is
\begin{subequations}
\label{Htwolayer}
\begin{align}
&{\cal H}_\pm=\begin{pmatrix}
H_\pm(\bm{k}-e\bm{A}+\bm{K})&\Delta_0 e^{i\phi}\\
\Delta_0 e^{-i\phi}&-H_\pm(\bm{k}+e\bm{A}-\bm{K})
\end{pmatrix},\\
&H_\pm(\bm{k})=\pm(v/a_0)\sigma_x \sin a_0 k_x\pm(v/a_0)\sigma_y \sin a_0 k_y\nonumber\\
&\qquad\qquad \pm \sigma_z M(k)-\mu,\\
&M(k)=M_0-(M_1/a_0^2)(2-\cos a_0 k_x-\cos a_0 k_y).
\end{align}
\end{subequations}
The Hamiltonian acts on a spinor with the four components
\begin{equation}
\Psi_\pm(\bm{k})=\frac{1}{\sqrt{2}}\begin{pmatrix}
[\psi_{\uparrow{\rm upper}}\pm\psi_{\uparrow{\rm lower}}](\bm{k})\\
[\psi_{\downarrow{\rm upper}}\mp\psi_{\downarrow{\rm lower}}](\bm{k})\\
-i[\psi_{\downarrow{\rm upper}}\pm\psi_{\downarrow{\rm lower}}]^\ast(-\bm{k})\\
i[\psi_{\uparrow{\rm upper}}\mp\psi_{\uparrow{\rm lower}}]^\ast(-\bm{k})
\end{pmatrix},
\end{equation}
for spin-up and spin-down electrons on the upper and lower surface of the topological insulator layer. The first two elements of the spinor $\Psi$ refer to electrons and the last two elements to holes. These are coupled by the \textit{s}-wave pair potential $\Delta_0$, which we take the same on both surfaces. The particle-hole symmetry relation is 
\begin{equation}
{\cal H}_\pm(\bm{k})=-\sigma_x\nu_y{\cal H}_\mp^\ast(-\bm{k})\sigma_x\nu_y,
\end{equation}
where the $\sigma_\alpha$ and $\tau_\alpha$ Pauli matrices act on the spin and electron-hole degree of freedom, respectively.

For the mass term $M(k)$ we take $M_0=0$, $M_1=0.2\,a_0v$, such that $H_0$ has a single gapless Dirac point at $\bm{k}=0$. Near this Dirac point the upper and lower surface are uncoupled, so the eigenstate can equivalently be written in the single-surface basis $(\psi_{\uparrow},\psi_{\downarrow},-i\psi_{\downarrow}^\ast,i\psi_{\uparrow}^\ast)$. The effect of a gap opening due to a nonzero $M_0$ is examined at the end of this Appendix.

The Hamiltonian is discretized on a square lattice (lattice constant $a_0$) with nearest neighbor hopping (hopping energy $v/a_0$). The magnetic field $B$ is uniform in the $z$-direction, vector potential $\bm{A}=-By\hat{x}$. The superflow momentum is $\bm{K}=K\hat{x}$. The amplitude $\Delta_0$ of the pair potential is taken as a constant, the phase $\phi(x,y)$ winds by $2\pi$ around each vortex.

We take a square vortex lattice, with lattice constant $d_0=Na_0$. The flux through each magnetic unit cell is $h/e$, so it contains a pair of $h/2e$ vortices. The integer $N$ determines the magnetic field via $B=(Na_0)^{-2}h/e$. The vortices are placed on the diagonal of the magnetic unit cell, at the positions $(x,y)=(N a_0/4)(1,1)$ and $(Na_0/4)(3,3)$. By taking for $N$ twice an odd integer, we ensure that the singularity in the phase field at the vortex core does not coincide with a lattice point. The phase field is discretized along the lines set out in App. B of Ref.\ \onlinecite{SPac18}. The eigenvalues and eigenfunctions of $H$ are calculated using the \textit{Kwant} tight-binding code \cite{Skwant}.

\subsection{Additional numerical results}
\label{sec_addition}

\begin{figure}[tb]
\centerline{\includegraphics[width=0.9\linewidth]{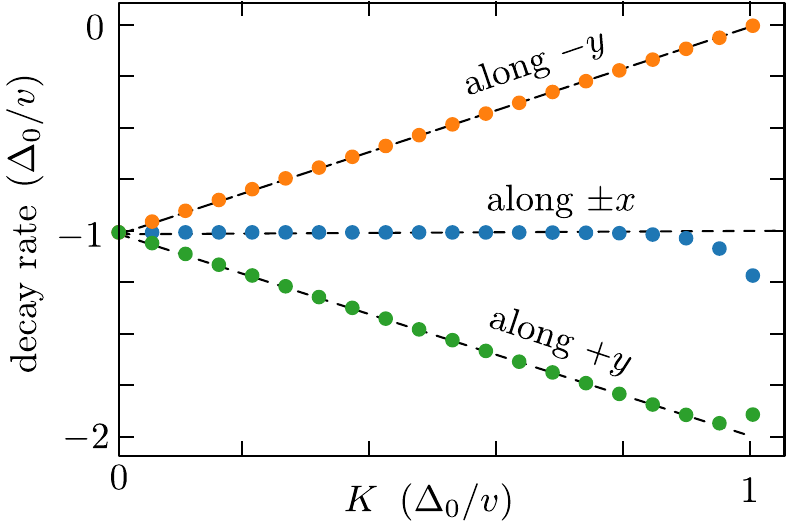}}
\caption{Decay rate of the Majorana mode confined to a vortex core. The data from the numerical simulation (colored points, $\Delta_0=20\, v/d_0$) closely follows the analytical prediction $|\Psi|\propto e^{-Ky}e^{-(\Delta_0/v)r}$ (dashed lines).
}
\label{fig_decayrate}
\end{figure}

Here we collect some additional results to those shown in the main text. In the confined phase $vK<\Delta_0$ we show in Fig.\ \ref{fig_decayrate} the anisotropic decay rates of the Majorana zero-modes bound to a vortex core, as in the left panel of Fig.\ \ref{fig_simulation1}. The localization length $(\Delta_0/v-K)^{-1}$ of the zero-modes diverges at the transition.

\begin{figure}[tb]
\centerline{\includegraphics[width=0.9\linewidth]{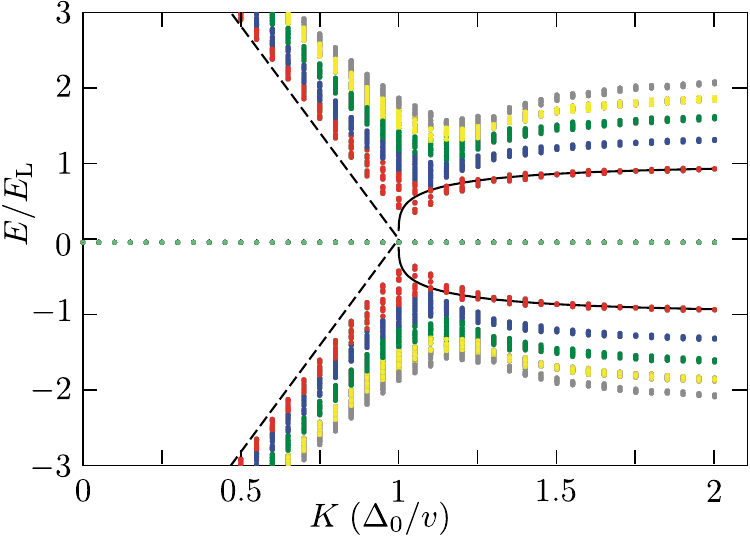}}
\caption{Excitation spectrum as a function of the superflow momentum (parameters as in Fig.\ \ref{fig_simulation1}). For $K<\Delta_0/v$ the states are confined to vortex cores and form a quasi-continuum, for $K>\Delta_0/v$ they are extended states arranged into a sequence of Landau levels (distinguished by different colors, the Majorana zero-modes are the light-green dots). The deconfinement transition at $K=\Delta_0/v$ is accompanied by a near closing of the gap to the first excited state. The dashed curves show the expected gap scaling $\propto (\Delta_0/v-K)$ and $\propto (K-\Delta_0/v)^{1/4}$ on the two sides of the transition.
}
\label{fig_gapclosing}
\end{figure}

Fig.\ \ref{fig_gapclosing} shows how at the deconfinement transition the quasi-continuum of excited states in the vortex core is reorganized into a sequence of Landau levels. The critical exponents for the gap closing are different on the two sides of the transition. In the confined phase the gap to the first excited state scales with the inverse localization length, so $\propto (\Delta_0/v-K)^{1}$. In the deconfined phase the gap scales with the Landau level separation $E_{\rm L}\propto\sqrt{q_{\rm eff}}$, so $\propto (K-\Delta_0/v)^{1/4}$.

\begin{figure}[tb]
\centerline{\includegraphics[width=1\linewidth]{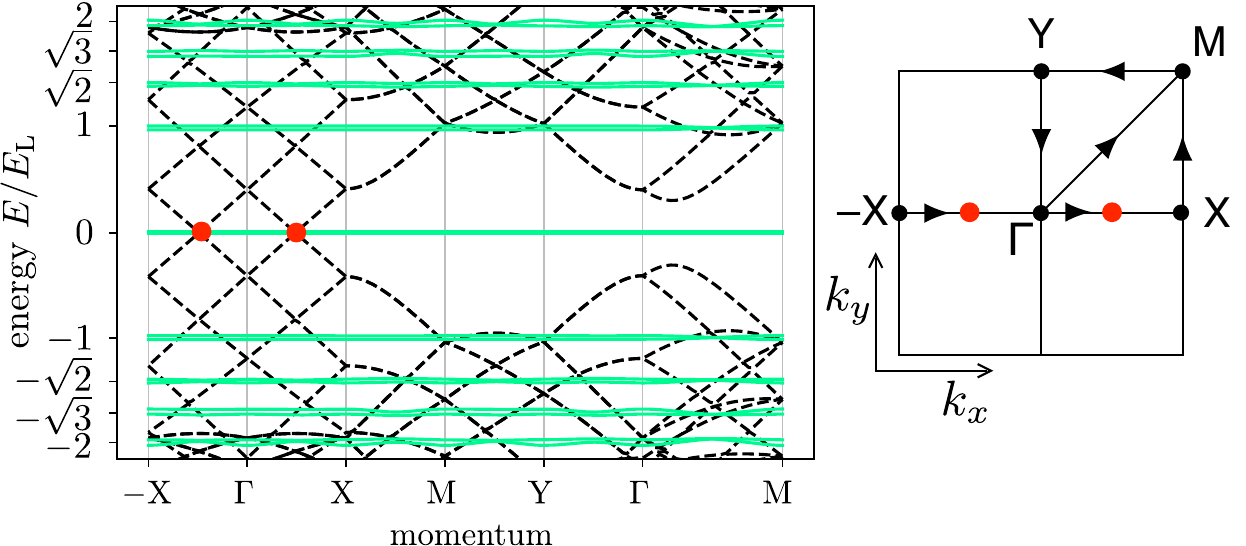}}
\caption{Dispersion relation in zero magnetic field (black dashed lines) and in the presence of the magnetic vortex lattice (green solid lines, the right panel shows the magnetic Brillouin zone). Both band structures are for $\mu=0$, and the same parameters as in Fig.\ \ref{fig_bandstructure}. The red dots indicate the Dirac points at $\bm{k}=(\pm Q,0)$ in zero magnetic field. The Landau levels are at $\pm\sqrt{n}\,E_{\rm L}$, $n=0,1, 2$, with $E_{\rm L}=\sqrt{4\pi q_{\rm eff}}\,\hbar v/d_0$.
}
\label{fig_fulldispersion}
\end{figure}

\begin{figure}[tb]
\centerline{\includegraphics[width=1\linewidth]{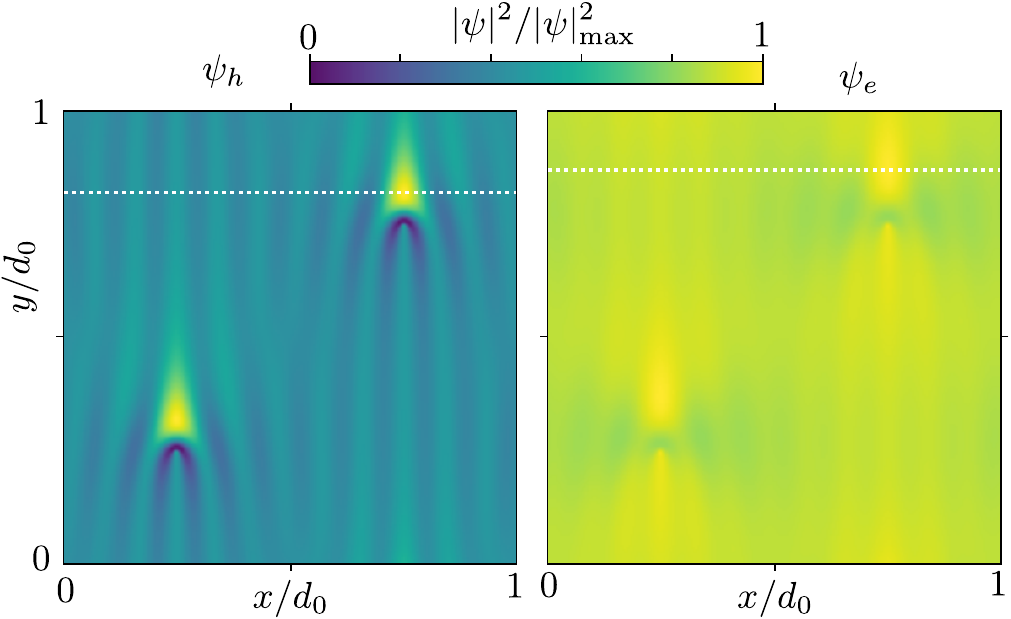}}
\caption{Local density of states in the unit cell of the vortex lattice, at the energy $E_0>0$ of the zeroth Landau level pushed above the Fermi level by a chemical potential $\mu>0$. The color scale plot shows $\sum_{\bm{k}}|\psi_{e,h}(x,y)|^2$, summed over the magnetic Brillouin zone, normalized to unit maximum value. The white dotted line indicates the cut shown in Fig.\ \ref{fig_DOS} of the main text, at the same parameters. The electron contribution to the local density of states (right panel) and the hole contribution (left panel) can be measured separately by tunnel spectroscopy at voltages $V=E_0$ and $V=-E_0$, respectively.
}
\label{fig_1square}
\end{figure}

In the deconfined phase $vK>\Delta_0$ we show in Fig.\ \ref{fig_fulldispersion} the Landau levels in the vortex lattice (complementing Fig.\ \ref{fig_bandstructure}). Fig.\ \ref{fig_1square} shows the local density of states in the zeroth Landau level. This shows the variation over the entire unit cell of the vortex lattice, to complement the line cut through a vortex core shown in Fig.\ \ref{fig_DOS} of the main text.

\subsection{Effect of overlap of top and bottom surface states}
\label{app_overlap}

\begin{figure}[tb]
\centerline{\includegraphics[width=0.8\linewidth]{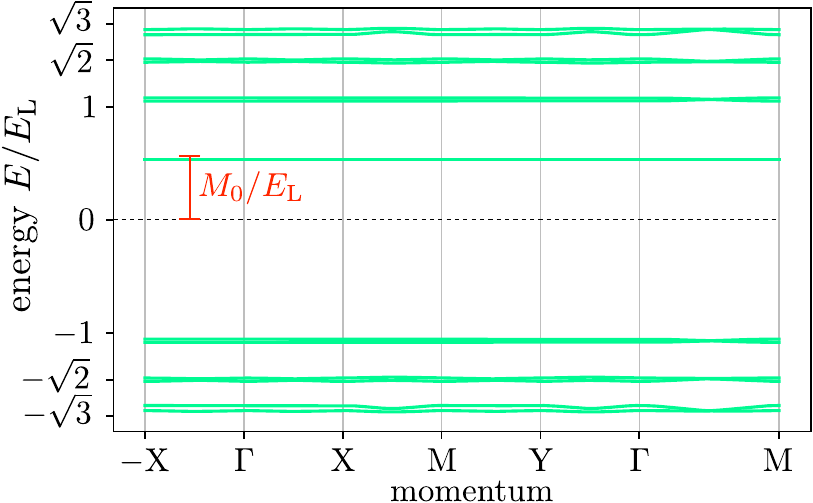}}
\caption{Same as Fig. \ref{fig_fulldispersion}, but now for a nonzero mass term $M_0$, to show how the hybridization gap shifts the zeroth Landau level away from $E=0$. The plot shows the spectrum of the Hamiltonian ${\cal H}_+$ in Eq.\ \eqref{Htwolayer}, the spectrum of ${\cal H}_-$ has the zeroth Landau level shifted to $-M_0$ (so that the full spectrum is particle-hole symmetric). The parameters are $K=2\Delta_0 =20\,\hbar v/d_0$, $d_0=102\,a_0$, $M_0=0.02/a_0$, $M_1=0.2\,a_0$.
}
\label{fig_overlap}
\end{figure}

A nonzero mass term $\pm M_0\sigma_z\nu_z$ in the Hamiltonian \eqref{Htwolayer} opens up a hybridization gap in the Dirac cone. Since the Majorana Landau level is an eigenstate of the chirality operator $\Lambda=\sigma_z\nu_z$, the effect of this term is to displace the flat band away from $E=0$ by an amount $M_0$. In Fig.\ \ref{fig_overlap} we show numerical results that demonstrate this. Provided that $M_0$ remains smaller than the Landau level separation $E_{\rm L}$, we do not expect the overlap of top and bottom surface states to prevent the detection of the Majorana Landau level. This is helpful because the overlap will favor a strong proximity effect on both surfaces.

\section{Solution of the Helmholtz equation for the Majorana Landau level}
\label{sec_solution}

The general solution of the 2D Helmholtz equation $\nabla^2 f=\Delta_0^2 f$ that governs the Majorana Landau level is a superposition of waves $e^{ipx\pm y\sqrt{p^2+\Delta_0^2}}$. Which superposition we need is determined by the requirement that $e^{-Ky-q(\bm{r})}f(x,y)$ is square integrable in the $x$--$y$ plane, with $K>\Delta_0>0$. We denote $Q=\sqrt{K^2-\Delta_0^2}$. For ease of notation we will set $\Delta_0\equiv 1$ in this appendix.

We construct a class of solutions for the case
\begin{align}
q(\bm{r})=\epsilon r-{\cal N}\ln \min(r,1),\;\;{\cal N}=1,2,\ldots,\label{qdev}
\end{align}
corresponding to $2{\cal N}$ vortices, each of vorticity $+2\pi$, at the origin. The positive infinitesimal $\epsilon>0$ is introduced to regularize integrals at $r\rightarrow\infty$. The restriction to an even number of overlapping vortices means that the branch cut which connects vortices pairwise can be ignored. (We have not succeeded in finding an analytical solution that incorporates the branch cut, but of course in the numerics this is not a limitation.)

The superposition of elementary solutions $e^{ipx\pm y\sqrt{p^2+1}}$ that cancels the exponential growth factor $e^{-Ky}$ has the general form
\begin{equation}
f=\begin{cases}
\int_{|p|>Q}dp\,C(p)e^{ipx+ y\sqrt{p^2+1}}&\text{if}\;\;y<0,\\
-\int_{|p|<Q}dp\,C(p)e^{ipx+ y\sqrt{p^2+1}}\\
\qquad+\int dp\,D(p)e^{ipx- y\sqrt{p^2+1}}&\text{if}\;\;y>0.
\end{cases}\label{fsolution}
\end{equation}
(We can use the symbol $C$ twice without loss of generality because the integration ranges do not overlap.)

The solution should be continuously differentiable at $r\neq 0$, which is satisfied if $f(x,y)$ and $\partial_y f(x,y)$ are continuous functions of $y$ at $y=0$, $x\neq 0$. The continuity requirement is that the Fourier transform $\int \cdots e^{ipx}dp$ of $C(p)$ equals the Fourier transform of $D(p)$ for $x\neq 0$, which means that $C(p)$ and $D(p)$ differ by a polynomial $L(p)$ of $p$. [Recall that the Fourier transform of a polynomial is given by derivatives of $\delta(x)$.] Similarly, the requirement of a continuous derivative is that $\sqrt{p^2+1}\,C(p)$ and $-\sqrt{p^2+1}\,D(p)$ differ by a polynomial $T(p)$. The unique solution of these two requirements is
\begin{equation}
\begin{split}
&C(p)=\frac{\tfrac{1}{2}T(p)}{\sqrt{p^2+1}}-\tfrac{1}{2}L(p),\\
&D(p)=\frac{\tfrac{1}{2}T(p)}{\sqrt{p^2+1}}+\tfrac{1}{2}L(p).
\end{split}\label{CDdef}
\end{equation}

We are free to choose a convenient basis for the polynomials $T(p)$ and $L(p)$, we will choose one for which the integral over $D(p)$ has a closed-form expression. The basis polynomials $T_n(p)$ and $L_n(p)$, $n=0,1,2,\ldots$ are
\begin{equation}
\begin{split}
&T_n(p)=\left(p+\sqrt{p^2+1}\right)^n+\left(p-\sqrt{p^2+1}\right)^n,\\
&L_n(p)=\frac{\left(p+\sqrt{p^2+1}\right)^n}{\sqrt{p^2+1}\,}-\frac{\left(p-\sqrt{p^2+1}\right)^n}{\sqrt{p^2+1}\,}.
\end{split}
\end{equation}
This choice of basis is related to a basis of Chebyshev polynomials ${\cal T}_n$, via the identities
\begin{equation}
\begin{split}
&T_n(p)=2(-i)^n{\cal T}_n(ip),\\
&L_n(p)=2 (-i)^{n-1} \sum _{m=0}^{n-1} {\cal T}_{2 m-n+1}(i p).
\end{split}
\end{equation}

Note that
\begin{equation}
T_{-n}(p)=(-1)^n T_n(p),\;\;L_{-n}(p)=-(-1)^nL_{-n}(p).
\end{equation}
A complete basis for the pairs of polynomials $T(p),L(p)$ is therefore given by the two sets $\{T_n,L_n\}\cup\{T_n,-L_n\}$ with $n=0,1,2,\ldots$, or equivalently by the single set $\{T_n,L_n\}$ with $n=0,\pm 1,\pm 2,\ldots$. The corresponding basis of the functions $C(p)$ and $D(p)$ in Eq.\ \eqref{CDdef} is
\begin{equation}
\begin{split}
&C_n(p)=\frac{\tfrac{1}{2}T_n(p)}{\sqrt{p^2+1}}-\tfrac{1}{2}L_n(p)=\frac{\left(p-\sqrt{p^2+1}\right)^n}{\sqrt{p^2+1}},\\
&D_n(p)=\frac{\tfrac{1}{2}T_n(p)}{\sqrt{p^2+1}}+\tfrac{1}{2}L_n(p)=\frac{\left(p+\sqrt{p^2+1}\right)^n}{\sqrt{p^2+1}},
\end{split}
\end{equation}
with $n=0,\pm 1,\pm 2,\ldots$.

We next use the Bessel function identities \cite{noteapp1}
\begin{equation}
\text{K}_n(r)=\begin{cases}
\tfrac{1}{2i^n}e^{in\theta}\int_{-\infty}^\infty dp\,D_n(p)e^{ipx- y\sqrt{p^2+1}} &\text{if}\;\;y\geq 0,\\
\tfrac{1}{2i^n}e^{in\theta}\int_{-\infty}^\infty dp\,C_n(p)e^{ipx+ y\sqrt{p^2+1}} & \text{if}\;\;y\leq 0,
\end{cases}\label{BesselKid}
\end{equation}
where $r=\sqrt{x^2+y^2}$ and $e^{i\theta}=(x+iy)/r$, to write the solution \eqref{fsolution} in the form
\begin{align}
f_n(x,y)=&-\int_{-Q}^{Q}dp\,\frac{\left(p-\sqrt{p^2+1}\right)^n}{\sqrt{p^2+1}}e^{ixp+y\sqrt{p^2+1}}\nonumber\\
&+ 2i^{n}e^{-in\theta}\text{K}_n( r),\label{fnresultApp}
\end{align}
which is Eq.\ \eqref{fnCndef} in the main text (upon restoring the units of $\Delta_0$).

The function $f_n$ is the first component of the spinor $\tilde{u}=(f,g)$, the second component is
\begin{equation}
 g_n=(i\partial_x-\partial_y)f_n=f_{n-1}.
\end{equation}

We now obtained an infinite countable set of solutions $\tilde{u}_n=(f_n,f_{n-1})$, $n=0,\pm 1,\pm 2,\ldots$ of the Helmholtz equation, such that $e^{-Ky}e^{-\epsilon r}\tilde{u}_n$ is square integrable at infinity. The condition that $r^{\cal N}\tilde{u}$ is square integrable at the origin (containing $2{\cal N}$ overlapping vortices) selects a finite subset. For $r\rightarrow 0$ we have $f_n\simeq r^{-|n|}$ if $n\neq 0$ and $f_0\simeq \ln r$. Normalizability requires that both $|n|\leq {\cal N}$ and $|n-1|\leq {\cal N}$, hence there are $2{\cal N}$ allowed values of $n\in\{-{\cal N}+1,-{\cal N}+2,\ldots {\cal N}-1,{\cal N}\}$.

All of this was for zero-modes $\Psi=(f,g,0,0)$ of positive chirality, in a lattice of $+2\pi$ vortices. Alternatively, we can consider zero-modes $\Psi=(0,0,f,g)$ of negative chirality in a lattice of $-2\pi$ vortices. The differential equations for $f$ and $g$ remain the same, but now the exponential factor that needs to be canceled is $e^{Ky}$ rather than $e^{-Ky}$. The sign change gives the negative chirality solution
\begin{subequations}
\begin{align}
f_n(x,y)={}&-\int_{-Q}^{Q}dp\,\frac{\left(p-\sqrt{p^2+1}\right)^n}{\sqrt{p^2+1}}e^{ixp-y\sqrt{p^2+1}}\nonumber\\
&+ 2i^{n}e^{in\theta}\text{K}_n( r),\\
g_n={}&(i\partial_x-\partial_y)f_n=-f_{n+1}.
\end{align}
\end{subequations}
The $2{\cal N}$ zero-modes are now labeled by the index $n\in\{-{\cal N},-{\cal N}+1,\ldots {\cal N}-2,{\cal N}-1\}$.

\section{Chain of vortices}
\label{sec_chain}

The regularization at infinity by the $\epsilon$ term in Eq.\ \eqref{qdev} is not needed if we have a periodic lattice of vortices. We demonstrate this by considering a linear chain of vortices at positions $\bm{R}_\ell$, spaced by $b$ at an angle $\vartheta\in[0,\pi/2]$ with the $x$-axis. We take a linear superposition of the solutions $e^{-Ky}f_n(\bm{r}-\bm{R}_\ell)$ from Eq.\ \eqref{fnresultApp}, with complex weights,
\begin{equation}
F_n(\bm{r})=\sum_{\ell=-\infty}^\infty e^{i\ell \kappa}e^{\ell K b\sin\vartheta}e^{-Ky}f_n(\bm{r}-\bm{R}_\ell).\label{Fnsuperposition}
\end{equation}
We do not include the envelope $e^{-q}$, because it tends to unity for large $r$ if we set $\epsilon\equiv 0$. The Bloch phase $\kappa$ is arbitrary.

\begin{widetext}
We substitute the large-$r$ expansion \eqref{largerexp},
\begin{equation}
F_n\rightarrow(-1)^n\sum_{\ell=-\infty}^\infty  e^{i\ell\kappa}\left(\frac{(K+Q)^ne^{-iQ(x-\ell b\cos\vartheta)}}{iK(x-\ell b\cos\vartheta) -Q(y-\ell b\sin\vartheta)}-\frac{(K-Q)^ne^{iQ(x-\ell b\cos\vartheta)}}{iK(x-\ell b\cos\vartheta)+Q(y-\ell b\sin\vartheta)}\right).\label{Fndef}
\end{equation}
We seek the decay of $F_n$ in the direction perpendicular to the chain, so for large $|\rho|$ when $(x,y)=(-\rho\sin\vartheta,\rho\cos\vartheta)$.
\end{widetext}

We thus need to evaluate an infinite sum of the form \cite{noteapp2}
\begin{subequations}
\label{Salphaz}
\begin{align}
S(\alpha,z)&=\sum_{\ell=-\infty}^\infty \frac{e^{i\ell\alpha}}{z+\ell },\;\;\alpha\in(0,2 \pi),\;\;z\in\mathbb{C}\backslash\mathbb{Z},\\
S(\alpha,z)&=\frac{2\pi  i}{e^{i\alpha z}-e^{i(\alpha-2\pi) z}}.\label{Salphazb}
\end{align}
\end{subequations}
In the limit $|{\rm Im}\,z|\rightarrow\infty$ this tends to
\begin{equation}
S(\alpha,z)\rightarrow\begin{cases}
-2\pi i e^{-(2\pi-\alpha){\rm Im}\,z}&\text{if}\;\; {\rm Im}\,z\rightarrow\infty,\\
2\pi i e^{\alpha {\rm Im}\,z}&\text{if}\;\; {\rm Im}\,z\rightarrow-\infty.
\end{cases}
\end{equation}

Substitution of Eq.\ \eqref{Salphaz} into Eq.\ \eqref{Fndef} gives, for $x=-\rho\sin\theta$, $y=\rho\cos\theta$,
\begin{align}
F_n\rightarrow{}& \frac{(-1)^n(K+Q)^n e^{iQ\rho\sin\vartheta} }{Qb\sin\vartheta-iKb\cos\vartheta}S\left(\alpha_+,z_-\right)\nonumber\\
&+\frac{(-1)^n(K-Q)^n e^{-iQ\rho\sin\vartheta} }{Qb\sin\vartheta+iKb\cos\vartheta}S\left(\alpha_-,z_+\right),
\end{align}
where we abbreviated
\begin{equation}
\begin{split}
&\alpha_\pm=\kappa\pm Qb\cos\vartheta\mod 2\pi,\\
&z_\pm=\frac{\rho}{b}\frac{\tfrac{1}{2}\sin 2 \vartheta\pm i KQ}{ K^2-\sin^2\vartheta}.
\end{split}
\end{equation}

Provided that $\alpha_\pm\neq 0\mod 2\pi$, the decay is exponential: $|F_n|\simeq e^{-c|\rho|/\lambda}$, with (reinserting the units of $\Delta_0$)
\begin{equation}
\lambda=b\frac{K^2-\Delta_0^2\sin^2\vartheta}{K\sqrt{K^2-\Delta_0^2}}
\end{equation}
and $c$ a coefficient of order unity that depends on the sign of $\rho$,
\begin{equation}
c=\begin{cases}
\min(\alpha_+,2\pi-\alpha_-)&\text{if}\;\;\rho>0,\\
\min(\alpha_-,2\pi-\alpha_+)&\text{if}\;\;\rho<0.
\end{cases}
\end{equation}
For a chain oriented along the $x$-axis or $y$-axis we have $\lambda$ equal to $bK/Q$ or $bQ/K$, respectively.

\section{Renormalized charge in the Majorana Landau level}
\label{sec_charge}

The charge expectation value of the deconfined zero-mode can be calculated by means of the block diagonalization approach of Ref.\ \onlinecite{SPac18}. Starting from the BdG Hamiltonian \eqref{Hdef} we first make the gauge transformation ${H}\mapsto U^\dagger {H}U$ with $U=\begin{pmatrix}
e^{i\phi}&0\\
0&1
\end{pmatrix}$, resulting in
\begin{align}
&H=\begin{pmatrix}
(\bm{k}+\bm{a}+\bm{q})\cdot\bm{\sigma}-\mu&\Delta_0\\
\Delta_0&-(\bm{k}+\bm{a}-\bm{q})\cdot\bm{\sigma}+\mu\end{pmatrix},\nonumber\\
&{\bm a}=\tfrac{1}{2}\nabla\phi,\;\;\bm{q}=\tfrac{1}{2}\nabla\phi-e\bm{A}+K\hat{x}.
\end{align}
We have included the chemical potential $\mu$.

For $K>\Delta_0$ in zero magnetic field there are gapless Dirac points at $\bm{k}=(k_x,k_y)=(\tilde{K},0)$ with 
\begin{equation}
\tilde{K}=\pm\kappa K,\;\;\kappa=\sqrt{1-\Delta_0^2/K^2}.
\end{equation}
To focus on the effect of a magnetic field on states near $\tilde{K}$ we set $k_x=\tilde{K}+\delta k_x$ and consider $\delta k_x$ small.

A unitary transformation ${H}\mapsto V^\dagger {H}V$ with
\begin{align}
&V=\begin{pmatrix}
\sigma_0\cos(\alpha/2)&\sigma_x\sin(\alpha/2)\\
-\sigma_x\sin(\alpha/2)&\sigma_0\cos(\alpha/2)
\end{pmatrix},\\
&\tan\alpha=-\Delta_0/\tilde{K},\;\;\cos\alpha=-(1+\Delta_0^2/\tilde{K}^2)^{-1/2}=-\kappa, \nonumber
\end{align}
approximately block-diagonalizes the Hamiltonian; the $2\times 2$ off-diagonal blocks contribute to the spectrum in second order in $\delta k_x$,  $\bm{a}$, $\bm{q}$, and $\mu$. The $2\times 2$ block along the diagonal that describes the hole-like states near $\bm{k}=(\kappa K,0)$ is given by
\begin{equation}
{H}_{+}=\kappa\mu-(\kappa\delta k_x+\kappa a_x-q_x)\sigma_x+(k_y+a_y-\kappa q_y)\sigma_y,
\end{equation}
while the electron-like states near $\bm{k}=(-\kappa K,0)$ are described by
\begin{equation}
{H}_{-}=-\kappa\mu+(\kappa\delta k_x+\kappa a_x+q_x)\sigma_x-(k_y+a_y+\kappa q_y)\sigma_y.
\end{equation}

The block diagonalization removes any interference between the electron and hole blocks, so this approximation cannot describe the striped density of states of Fig.\ \ref{fig_simulation1} --- for that we need the Helmholtz equation considered in the main text. Because the charge operator $\hat{Q}=-e\partial H_\pm/\partial\mu=\mp\kappa e$ commutes with $H_\pm$, the expectation value is given simply by
\begin{equation}
\langle \hat{Q}\rangle=\mp\kappa e\Rightarrow q_{\rm eff}=\kappa.
\end{equation}

The Fermi velocity in the $x$-direction is renormalized by the same factor, $v_x=\kappa v$, while $v_y$ is unaffected. This affects the Landau level energy $E_{\rm L}=\sqrt{4\pi}\,\hbar v_{\rm eff}/d_0$ of the anisotropic Dirac cone, via $v_{\rm eff}=\sqrt{v_xv_y}=\sqrt{\kappa} v$.

\section{Comparison of numerics and analytics}
\label{sec_comparison}

\begin{figure}[tb]
\centerline{\includegraphics[width=1\linewidth]{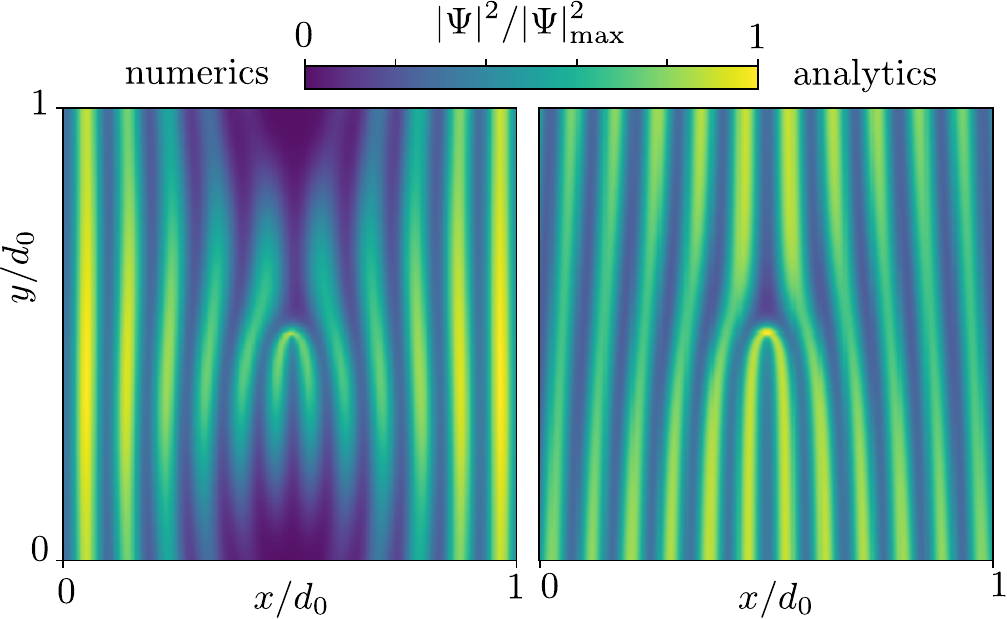}}
\caption{Comparison between numerical and analytical intensity profiles $|\Psi(x,y)|^2$, normalized to unit maximal value, for one of the two reflection-symmetric states in the zeroth Landau level. The parameter values are the same as in Fig.\ \ref{fig_comparison}, which compared the other state.
}
\label{fig_comparison2}
\end{figure}

In order to compare the analytic solution \eqref{fnCndef} of the Helmholtz equation with the numerical results from the tight-binding Hamiltonian \eqref{H02x2} we proceed as follows. For the analytic solution we take a single pair of vortices located at $\bm r = 0$, in a uniform magnetic field with total flux $h/e$ in a large disc centered at the origin. There are then two independent zero-modes $u$, $u'$ given by Eq.\ \eqref{uanalytic} with $q(r) = -\ln r$.

For the numerical calculation we consider an infinite lattice of vortices, with pairs of vortices positioned at points $\bm R_{\bm n} = d_0 \bm n$, $\bm n \in \mathbb{Z}^2$, in a uniform magnetic field $B=(h/e)d_0^{-2}$, vector potential $\bm{A}=-B(y,0)$. The Hamiltonian commutes with the magnetic translation operator
\begin{equation}
\begin{split}
&\mathcal{T}_{\bm n} = \begin{pmatrix}
e^{i h n_y x/d_0} &0\\
0& e^{-i h n_y x/d_0}
\end{pmatrix}T_{\bm n} \,, \\
& T_{\bm n}\bm r T_{\bm n}^\dag = \bm r + d_0\bm n\,.
\end{split}
\end{equation}
(The $2\times 2$ matrix acts on the electron-hole degree of freedom.) The eigenvalue $e^{i \bm k\cdot \bm n}$ of the eigenstates defines the magnetic momentum $\bm k\in [0,2\pi)^2$. At each value of $\bm k$ there are two independent zero-modes.

To make sure we are comparing the same state in the degenerate manifold we consider the operator product
\begin{equation}
\mathcal{P}_x = 
\begin{pmatrix}
0& e^{\frac{1}{2}i\phi(\bm{r})} \\ e^{-\frac{1}{2}i\phi(\bm{r})}&0
\end{pmatrix}
\sigma_x P_x
\begin{pmatrix}
e^{-\frac{1}{2}i\phi(\bm{r})} &0\\
0& e^{\frac{1}{2}i\phi(\bm{r})}
\end{pmatrix},
\end{equation}
with eigenvalues $\pm 1$, which is a symmetry respected both by the analytic and by the numerical calculation. The operator $P_x$ is the mirror symmetry operator in the $x$-direction,
\begin{equation}
P_x x P_x^\dag = -x \,,\quad
P_x y P_x^\dag = y\, .
\,
\end{equation}
The magnetic momentum transforms under $\mathcal{P}_x$ as $k_x\mapsto -k_x$, $k_y\mapsto k_y$. 

For the comparison we set $\bm k =0$, which is invariant under the action of $\mathcal{P}_x$. Then we can take the two zero-modes obtained numerically to be eigenstates of $\mathcal{P}_x$, and compare them with the corresponding eigenstates obtained analytically. Those are
\begin{equation}
u_{\pm}(\bm r) = u(\bm r) \pm u'(\bm r)\,,
\end{equation}
which, in view of the fact that
\begin{equation}
f_n(-x,y) = f_n^*(x,y)
\end{equation}
are eigenfunctions of ${\cal P}_x$ with eigenvalues $\pm 1$. Figs.\ \ref{fig_comparison} and \ref{fig_comparison2} compare the modulus squared of the $+1$ and $-1$ eigenstates of $\mathcal{P}_x$ respectively, with quite satisfactory correspondence.


\newpage

\begin{thebibliography}{99}
\bibitem{Vol07} G. E. Volovik, \textit{Quantum phase transitions from topology in momentum space}, Lect. Notes Phys. \textbf{718}, 31 (2007).
\bibitem{Agt17} D. F. Agterberg, P. M. R. Brydon, and C. Timm, \textit{Bogoliubov Fermi surfaces in superconductors with broken time-reversal symmetry}, Phys. Rev. Lett. \textbf{118}, 127001 (2017).
\bibitem{Yua18} Noah F. Q. Yuan and Liang Fu, \textit{Zeeman-induced gapless superconductivity with a partial Fermi surface}, Phys. Rev. B \textbf{97}, 115139 (2018).
\bibitem{Aut20} S. Autti, J. T. M\"{a}kinen, J. Rysti, G. E. Volovik, V. V. Zavjalov, and V. B. Eltsov, \textit{Exceeding the Landau speed limit with topological Bogoliubov Fermi surfaces}, Phys. Rev. Res. \textbf{2}, 033013 (2020).
\bibitem{Lin20} J. M. Link and I. F. Herbut, \textit{Bogoliubov-Fermi surfaces in non-centrosymmetric multi-component superconductors}, Phys. Rev. Lett. \textbf{125}, 237004 (2020).
\bibitem{Zhu20} Zhen Zhu, Micha{\l} Papaj, Xiao-Ang Nie, Hao-Ke Xu, Yi-Sheng Gu, Xu Yang, Dandan Guan, Shiyong Wang, Yaoyi Li, Canhua Liu, Jianlin Luo, Zhu-An Xu, Hao Zheng, Liang Fu, and Jin-Feng Jia, \textit{Discovery of segmented Fermi surface induced by Cooper pair momentum}, arXiv:2010.02216. For a commentary on this experiment, see DOI: 10.36471\/JCCM\_October\_2020\_01
\bibitem{Rac16} S. Rachel, L. Fritz, and M. Vojta, \textit{Landau levels of Majorana fermions in a spin liquid}, Phys. Rev. Lett. \textbf{116}, 167201 (2016).
\bibitem{Per17} B. Perreault, S. Rachel, F. J. Burnell, and J. Knolle, \textit{Majorana Landau-level Raman spectroscopy}, Phys. Rev. B \textbf{95}, 184429 (2017).
\bibitem{Fu08} Liang Fu and C. L. Kane, \textit{Superconducting proximity effect and Majorana fermions at the surface of a topological insulator}, Phys. Rev. Lett. \textbf{100}, 096407 (2008).
\bibitem{note6} The term $K\sigma_x$ in the BdG Hamiltonian \eqref{Hdef} is equivalent, upon a gauge transformation, to a gradient $2Kx$ in $\phi$.
\bibitem{note11} The overlap of states on the top and bottom surfaces of the topological insulator thin film shifts the Majorana Landau away from $E=0$ by the hybridization gap, while keeping the spatial structure of the wave functions intact. We include this effect in the calculations in App.\ \ref{sec_numerics} of the Supplemental Material.
\bibitem{Jac81} R. Jackiw and P. Rossi, \textit{Zero modes of the vortex-fermion system}, Nucl. Phys. B \textbf{190}, 681 (1981).
\bibitem{note1} To understand how the solution \eqref{psismallpssolution} relates to the ${K}=0$ solution in Ref.\ \cite{Jac81}, note the (non-unitary) transformation $e^{{K}y\Lambda}H_\pm e^{{K}y\Lambda}=H_\pm+{K}\sigma_x$, with $\Lambda={\rm diag}\,(1,-1,-1,1)$. The spinor $\Psi_\pm$ is an eigenstate of $\Lambda$ with eigenvalue $\pm 1$, so if $H_\pm \Psi_\pm=0$ for ${K}=0$, then $H_\pm e^{\pm {K}y}\Psi_\pm=0$ for ${K}\neq 0$.
\bibitem{data1} The data in Fig.\ \ref{fig_simulation1} is obtained from the tight-binding Hamiltonian \eqref{H02x2} of the topological insulator layer. The parameters are $\Delta_0=20\, \hbar v/d_0$, $d_0=302\,a_0$, $B=h/ed_0^2$, $\mu=0$, $M_0=0$, $M_1=0.2\,a_0$. The vortex pair in a unit cell is at the positions $(x,y)=(d_0/4)(1,1)$ and $(d_0/4)(3,3)$. The superflow momentum $K$ equals $0.8\,\Delta_0/v$ in the left panel and $2\,\Delta_0/v$ in the right panel.
\bibitem{note9} The anisotropic decay of the Majorana zero-mode in the left panel of Fig.\ \ref{fig_simulation1} can be understood as the effect of the Magnus force which the superflow momentum $\bm{K}=K\hat{x}$ exerts on the axial spin $\bm{S}={\cal C}\hat{z}$ of the Majorana fermions (as determined by their chirality ${\cal C}=\pm 1$). The direction of slow decay of the zero-mode is given by the cross product ${\bm K}\times\bm{S}$.
\bibitem{Kat08} M. I. Katsnelson and M. F. Prokhorova, \textit{Zero-energy states in corrugated bilayer graphene}, Phys. Rev. B \textbf{77}, 205424 (2008).
\bibitem{Kai09} J. Kailasvuori, \textit{Pedestrian index theorem \`{a} la Aharonov-Casher for bulk threshold modes in corrugated multilayer graphene}, EPL \textbf{87}, 47008 (2009).
\bibitem{Pac18} M. J. Pacholski, C. W. J. Beenakker, and I. Adagideli, \textit{Topologically protected Landau level in the vortex lattice of a Weyl superconductor}, Phys. Rev. Lett. \textbf{121}, 037701 (2018).
\bibitem{Aha79} Y. Aharonov and A. Casher, \textit{Ground state of a spin-1/2 charged particle in a two-dimensional magnetic field}, Phys. Rev. A \textbf{19}, 2461 (1979).
\bibitem{Mel99} A. S. Mel'nikov, \textit{Quantization of the quasiparticle spectrum in the mixed state of d-wave superconductors}, J. Phys. Condens. Matter \textbf{11}, 4219 (1999).
\bibitem{Fra00} M. Franz and Z. Te\v{s}anovi\'{c}, \textit{Quasiparticles in the vortex lattice of unconventional superconductors: Bloch waves or Landau levels?}, Phys. Rev. Lett. \textbf{84}, 554 (2000).
\bibitem{note2} The integral equation \eqref{qintegral} for $q(\bm{r})$ follows from the definition \eqref{qdef}, which implies that $\nabla^2 q(\bm{r})=\hat{z}\cdot\nabla\times(e\bm{A}-\tfrac{1}{2}\nabla\phi)=eB-\pi\sum_{n}\delta(\bm{r}-\bm{R}_n)$. The Green function of this 2D Poisson equation is $(2\pi)^{-1}\ln|\bm{r}-\bm{r}'|$. Also note that $\Phi_0\equiv \pi/e$ in units where $\hbar\equiv 1$.
\bibitem{note3} We assume there is an even number of vortices in $S$. If the number of vortices is odd, a zero-energy edge state along the perimeter of $S$ will ensure that the total number of Majorana zero-modes remains even.
\bibitem{note4} This normalization requirement at the vortex core ties the chirality of the Majorana zero-modes to the sign of the vorticity. If we would have chosen $-2\pi$ vortices the field $q(\bm{r})$ would tend to $+\tfrac{1}{2}\ln|r-\bm{R}_n|$ near a vortex core, and the product $e^{-q}f\propto |\bm{r}-\bm{R}_n|^{-1/2}f $ would not have been square integrable.
\bibitem{Appsolution} Details of the solution of the Helmholtz equation are given in Apps. \ref{sec_solution} and \ref{sec_chain} of the Supplemental Material.
\bibitem{note10} The $1/r$ decay of the deconfined Majorana zero-mode implies a density of states peak which decays slowly $\propto 1/\ln L$ as a function of the system size $L$. There is a formal similarity here with the zero-modes originating from vacancies in a 2D bipartite lattice \cite{Sut86,Per06}.
\bibitem{Sut86} B. Sutherland, \textit{Localization of electronic wave functions due to local topology}, Phys. Rev. B \textbf{34}, 5208 (1986).
\bibitem{Per06} V. M. Pereira, F. Guinea, J. M. B. Lopes-dos Santos, N. M. R. Peres, and A. H. Castro-Neto, \textit{Disorder induced localized states in graphene}, Phys. Rev. Lett. \textbf{96}, 036801 (2006).
\bibitem{Sha10} Wen-Yu Shan, Hai-Zhou Lu, and Shun-Qing Shen, \textit{Effective continuous model for surface states and thin films of three-dimensional topological insulators}, New J. Phys. \textbf{12}, 043048 (2010).
\bibitem{Zha15} Song-Bo Zhang, Hai-Zhou Lu, and Shun-Qing Shen, \textit{Edge states and integer quantum Hall effect in topological insulator thin films}, Scientif. Rep. \textbf{5}, 13277 (2015).
\bibitem{note5} In the basis $\Psi=(\psi_{\uparrow{\rm upper}},\psi_{\downarrow{\rm upper}},\psi_{\uparrow{\rm lower}},\psi_{\downarrow{\rm lower}})$ the $4\times 4$ Hamiltonian of the topological insulator layer is $H_0=t_0\textstyle{\sum_{j=x,y}} \tau_z\sigma_j \sin k_j a_0+\tau_x \sigma_0 M(k)-\mu$, with Pauli matrix $\tau_z$ acting on the layer index. A unitary transformation block-diagonalizes the Hamiltonian. One of the $2\times 2$ blocks is given in Eq.\ \eqref{H02x2}, the other block has $M$ replaced by $-M$.
\bibitem{kwant} C. W. Groth, M. Wimmer, A. R. Akhmerov, and X. Waintal, \textit{Kwant: A software package for quantum transport}, New J. Phys. \textbf{16}, 063065 (2014).
\bibitem{Apsimulation} Details of the method of numerical simulation, with supporting data, are given in App.\ \ref{sec_numerics} of the Supplemental Material.
\bibitem{appQeff} The renormalized charge $q_{\rm eff}$ in the Majorana Landau level is calculated in App.\ \ref{sec_charge} of the Supplemental Material. That calculation also gives the renormalized Fermi velocity $v_{\rm eff}=\sqrt{v_xvy}=\sqrt{q_{\rm eff}}\,v$ that appears in the Landau level energy $E_{\rm L}$.
\bibitem{note7} The chiral symmetry at $\mu=0$ is broken by the mass term $M(k)$ in the Hamiltonian \eqref{H02x2}. This residual chiral symmetry breaking is visible in Fig. \ref{fig_bandstructure} as a very small splitting of the $\mu=0$ Landau levels (green flat bands).
\bibitem{note8} The comparison between numerics and analytics in Fig.\ \ref{fig_comparison} involves no adjustable parameters. To compare the same state in the degenerate zeroth Landau level we choose the state with left-right reflection symmetry. There are two of these, the other is compared in App.\ \ref{sec_comparison} of the Supplemental Material.
\bibitem{Sun17} Hao-Hua Sun and Jin-Feng Jia, \textit{Detection of Majorana zero mode in the vortex}, npj Quantum Mat. \textbf{2}, 34 (2017). 
\bibitem{Cho13} S. Cho, B. Dellabetta, A. Yang, J. Schneeloch, Z. Xu, T. Valla, G. Gu, M. J. Gilbert, and N. Mason, \textit{Symmetry protected Josephson supercurrents in three-dimensional topological insulators}, Nature Comm. \textbf{4}, 1689 (2013). 
\bibitem{Pik17} D. I. Pikulin and M. Franz, \textit{Black hole on a chip: proposal for a physical realization of the SYK model in a solid-state system}, Phys. Rev. X \textbf{7}, 031006 (2017).
\end{thebibliography}

\begin{thebibliography}{99}
\makeatletter
\makeatother
\bibitem{SPac18} M. J. Pacholski, C. W. J. Beenakker, and I. Adagideli, \textit{Topologically protected Landau level in the vortex lattice of a Weyl superconductor}, Phys. Rev. Lett. \textbf{121}, 037701 (2018).
\bibitem{Skwant} C. W. Groth, M. Wimmer, A. R. Akhmerov, and X. Waintal, \textit{Kwant: A software package for quantum transport}, New J. Phys. \textbf{16}, 063065 (2014).
\bibitem{noteapp1} The identities \eqref{BesselKid} follow from the integral representation 
$\text{K}_n(r)=\tfrac{1}{2}(r/2)^n\int_0^\infty t^{-n-1}\exp(-t-\tfrac{1}{4}r^2/t)\,dt$, upon the substitution $p=\tfrac{1}{2}(t-1/t)$.
\bibitem{noteapp2} For a derivation of Eq. \eqref{Salphazb}, and its relation to the Lerch zeta function, see\\ 
\url{https://mathoverflow.net/q/379157/11260}.

\end{thebibliography}
\end{document}